\documentclass[aip, rsi, reprint]{revtex4-1}

\usepackage{graphicx, xcolor}
\usepackage{amsmath}
\usepackage[hidelinks]{hyperref}
\usepackage{epsfig}
\usepackage{nicefrac}
\usepackage{textcomp}
\usepackage{natbib}
\usepackage{siunitx}
\newcommand{\ket}[1]{|#1\rangle}

\begin{document}

\widetext

\title{Nuclear Quantum-Assisted Magnetometer} 
\vskip 0.25cm
\author{Thomas H\"aberle}
\author{Thomas Oeckinghaus}
\author{Dominik Schmid-Lorch}
\author{Matthias Pfender}
\author{Felipe F\'avaro de Oliveira}
\author{Seyed Ali Momenzadeh}
\author{Amit Finkler}
\email[{Author to whom correspondence should be addressed.
   Email:~}]{a.finkler@physik.uni-stuttgart.de}
\affiliation{3.\ Physikalisches Institut, Universit\"at Stuttgart, Pfaffenwaldring 57, 70569 Stuttgart, Germany}
\author{J\"org Wrachtrup}
\affiliation{3.\ Physikalisches Institut, Universit\"at Stuttgart, Pfaffenwaldring 57, 70569 Stuttgart, Germany}
\affiliation{Max Planck Institute for Solid State Research, Heisenbergstr.\ 1, 70569 Stuttgart, Germany}

\date{\today}

\begin{abstract}
Magnetic sensing and imaging instruments are important tools in biological and material sciences.
There is an increasing demand for attaining higher sensitivity and spatial resolution, with implementations using a single qubit offering potential improvements in both directions.
In this article we describe a scanning magnetometer based on the nitrogen-vacancy center in diamond as the sensor. 
By means of a quantum-assisted readout scheme together with advances in photon collection efficiency, our device exhibits an enhancement in signal to noise ratio of close to an order of magnitude compared to the standard fluorescence readout of the nitrogen-vacancy center.
This is demonstrated by comparing non-assisted and assisted methods in a $T_1$ relaxation time measurement.
\end{abstract}

\maketitle

\section{Introduction}

Spin defects in diamond are novel nanoscale sensors.
While the size of a single defect guarantees nanometer spatial resolution in scanning probe arrangements, reading out the signal of single defects may significantly increase acquisition times.

The quality of the experimental readout is measured by its signal to noise ratio (SNR).
Since the SNR is primarily photon shot-noise limited, it can be improved by an increased fluorescence collection efficiency.
With quantum-assisted readout schemes, however, it is possible to enhance the SNR even further.
In this paper we present a quantum metrology and imaging setup, which was optimized for achieving the highest SNR by employing such a quantum-assisted readout scheme in combination with improved collection efficiency.
We use the negatively-charged nitrogen-vacancy (NV) center in diamond, which is nowadays one of the most prominent candidates for quantum applications, especially at ambient conditions.\cite{Jelezko2006}

At a finite magnetic field, the NV can be considered as a single qubit, which can be initialized and read-out optically, while its spin state can be manipulated with the use of microwave-frequency irradiation.
Its key signature is the ability to detect its spin-state by measuring its fluorescence intensity - further on abbreviated as optically detected magnetic resonance (ODMR).\cite{Gruber1997}
For a detailed review of the NV center properties, we refer the reader to Ref.\,\onlinecite{Doherty2013}.

NV ensembles reach magnetic field sensitivities as high as \si[per-mode=symbol]{\pico\tesla}$/\sqrt{\mathrm{Hz}}$, \cite{Wolf2014} and single NV centers have been demonstrated to sense a single electron spin at a distance of \SI{50}{\nano\meter} (see Ref.~\onlinecite{Grinolds2013}).
Even the detection of single nuclear spins seems to be within reach.\cite{Mueller2014, Sushkov2014b}
Moreover, single defect center scanning probe microscopy has reached resolutions in the \SIrange[range-phrase=-, range-units=single]{10}{20}{\nano\meter} range.\cite{Chernobrod2005, Degen2008, Maze2008, Balasubramanian2008, Taylor2008, Thiel2016, HaeberleT.2015, RugarD.2014, DeVience2015}

Despite such remarkable applications, long integration times are still necessary to reach satisfactory SNRs.
In spectroscopy applications it is possible to run measurements over several days, since probe and sample form a mechanically static system.
However, such readout methods are quite tenuous, inefficient and prevent the observation of even slow dynamics.
This is especially critical in the case of scanning probe applications, where it becomes difficult to achieve long-term mechanical stability since the sample and probe are scanned with respect to each other. 
Additionally, since the measurement has to be repeated for each pixel, frame acquisition times increase quadratically with increasing pixel density, limiting the resolution for any given scan range.

It is thus of paramount importance to improve SNR.
Various ideas ranging from enhanced photon collection efficiencies by nano-engineered waveguides\cite{Hausmann2010, Momenzadeh2015} to an improvement of the readout fidelity of the spin state itself were explored.
The latter approach was demonstrated e.g.\ via charge state conversion\cite{Shields_2015} or nuclear spin-assisted readout schemes such as repetitive readout\cite{Jiang2009} and single shot readout (SSR).\cite{PhysRevB.81.035205, Neumann2010b}
Here, we chose to pursue SSR to improve the readout fidelity, which also allows the use of more sophisticated schemes like quantum memories or error correction.

This article consists of three main sections.
First, in Sec.~\ref{improvements}, we shed light on how SNR is improved via SSR.
Then, in Sec.~\ref{setup} we illustrate the different enhancements and modifications we have made to a commercial platform in order to achieve this SNR enhancement.
Finally, in Sec.~\ref{performance}, we present results exhibiting a gain of 8.6 in SNR.

\section{Nuclear Quantum-Assisted Electron Spin Readout}\label{SSR}\label{improvements}

NV center based measurements start with the initialization of the spin followed by radio frequency control fields to render the NV center's electron spin sensitive to a certain physical quantity.
This is then followed by the actual spin readout.

Conventionally, the NV center is read out optically by applying a laser pulse.
However, the resulting fluorescence typically yields less than 0.1 significant photons per readout.
Thus, since the spin state is destroyed during readout, the whole measurement scheme has to be repeated several hundred thousand times to build up enough statistics.
Since a single measurement run can easily take several milliseconds, this renders the measurement scheme rather inefficient.

Consequently, several approaches have been developed to circumvent this problem.
The general idea is to use a different spin readout mechanism which does not destroy the spin state, a so-called quantum non-demolition (QND) measurement. \cite{QND}
The readout of a single measurement run can then be repeated several times, thus enhancing the SNR.

For example, at low temperature ($T<\SI{20}{\kelvin}$, see Ref.~\onlinecite{Goldman2015}) it is possible to resonantly excite the different sub-levels of the NV center to read out the spin state quite efficiently. \cite{Robledo2011}
At ambient conditions, however, it is necessary to use auxiliary memories with longer life time.
One way to implement this is to read out the state of the NV center via spin-to-charge conversion. \cite{Shields_2015, Hopper2016}
Spin state dependent ionization and subsequent charge state dependent readout allows measurement of the spin state to a certainty only a factor of two away from the spin projection noise level.\cite{Shields_2015}
Technically, however, this approach is rather challenging, since the use of lasers with three different wavelengths becomes necessary to perform charge conversion and readout.

Another alternative is to swap the spin state to a second spin which can be read out with a QND measurement.
Here, the number of readout cycles is ultimately limited by the lifetime of the second spin.
Therefore, nuclear spins are prime candidates, since they feature longer lifetimes than electron spins.
The NV center has a built-in memory in the form of the nitrogen nucleus, which has a nuclear spin of either $I=1$ for $^{14}\mathrm{N}$ or $I=1/2$ for $^{15}\mathrm{N}$.
Another candidate would be a $^{13}\mathrm{C}$ isotope ($I = 1/2$) in the diamond lattice.
However, the proportion of NV centers with a $^{13}\mathrm{C}$ nuclear spin at a suitable position is not very high.

Two variations of nuclear spin-enhanced readout were demonstrated in the past.
The so-called repetitive readout\cite{Jiang2009} and the single shot readout (SSR) \cite{Neumann2010b} employed in this work.
It is important to note that both methods are projection measurements,
i.e.\ the measurement still needs to be repeated to determine the actual spin state.
However, instead of several hundred of thousand repetitions, only a few thousand repetitions suffice.

As a figure of merit for the quality of the SSR scheme we use the nuclear spin readout fidelity, $\mathcal{F}$, which reflects the probability, that two subsequent nuclear spin readouts yield the same result.\footnote{This definition slightly differs from previously used definitions,\cite{Waldherr_2011} but is more straightforward for our purposes.}
There are two reasons for this not to be the case: Either the nuclear spin flipped of its own accord, or the measurement of the nuclear spin state yielded the wrong result.
The probability for a random nuclear spin flip depends on the nuclear spin lifetime, and is therefore less likely for short readout times.
The correct readout of the nuclear spin state, however, improves with the number of collected photons, which can be increased by a longer readout time.
Thus we have two competing mechanisms leading to an optimal readout time with the highest SSR fidelity (see section~\ref{performance_SSR}).
One needs to keep in mind, that the aforementioned nuclear spin lifetime is the lifetime under laser illumination of the NV center and should not be confused with the nuclear spin lifetime in the dark, which is considerably longer. \cite{Neumann2010b, Pfender2016}

In order to improve the lifetime of the nuclear spin it needs to be decoupled from the electron spin.
It is therefor necessary to detune the NV center as far as possible from the excited state level anticrossing (at $\sim\SI{50}{\milli\tesla}$), by applying a strong magnetic bias field.\cite{Neumann2010b}
However, the increase in nuclear spin lifetime only prevails, if the applied magnetic field is well aligned along the axis of the NV center.
Reaching the necessary quality of alignment and stability of the magnetic field poses the true challenge if one wants to establish SSR.
The design of the magnet and its thermal stabilization are described in full detail in section~\ref{setup}.

\begin{figure}[!htbp]
		\centering
		\includegraphics[width=1.0\columnwidth]{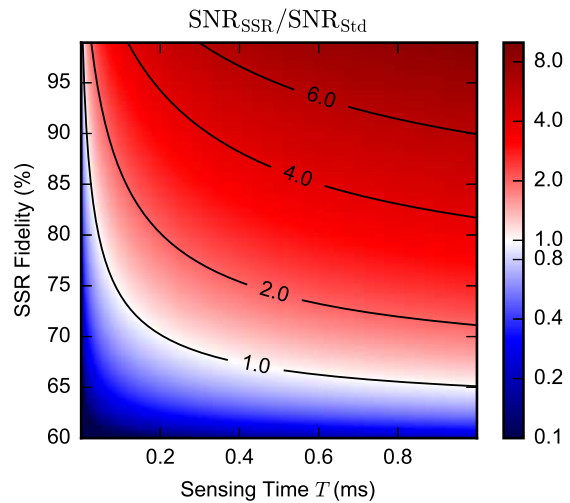}
  \caption{Theoretically predicted improvement factor of the SSR-assisted SNR vs.\ the standard SNR (Std), $\mathrm{SNR}_\mathrm{SSR}/\mathrm{SNR}_\mathrm{Std}$, as a function of SSR fidelity and sensing time $T$. Isolines of the improvement factor are displayed as guides to the eye. All parameters that are not varied here were taken from the real experiment.}
	\label{fig:field}
\end{figure}

The potential improvement of the SNR using SSR was estimated from theory and is displayed in Fig.\,\ref{fig:field} (full detail derivation is given in section~\ref{appendices}).
Figure~\ref{fig:field} shows the relative improvement in SNR between a SSR and a non-SSR measurement as a function of initial SSR fidelity and sensing time.
Since SSR has a non-trivial overhead compared to the standard readout, it is less efficient for short sensing times but thrives for longer sensing times.
From the simulation it becomes apparent, that with an SSR fidelity of e.g. \SI{90}{\percent}, the break-even-point would already be reached for a sensing time as short as \SI{30}{\micro\second}.

As shown in section~\ref{performance}, we actually achieve SSR fidelity higher than \SI{90}{\percent}.
However, to reach this goal we further need to enhance the photon collection efficiency in addition to improving the magnetic field.

This feat was achieved by nano-engineering the diamond sample into efficient waveguides in the form of tapered nanopillars. \cite{Hausmann2010, Momenzadeh2015}
This way, the collection efficiency is typically improved by a factor of up to ten or more compared to unstructured diamond.
Therefore, this improvement alone boosts the SNR by a factor of about three.
Combined with SSR, however, it allows us to reduce the number of readout cycles, which effectively reduces the need for overly long nuclear spin lifetimes and this finally gives us the high fidelity we are aiming for.

\section{System Design}\label{setup}

The room temperature microscope is built around a confocal and scanning force microscope (attoCSFM), combining a confocal fluorescence microscope (CFM) and an atomic force microscope (AFM) from \mbox{attocube~systems~AG}.\footnote{\href{http://www.attocube.com/attomicroscopy/microscopy-solutions/odmr/}{Attocube - combined atomic force \& confocal microscope}}
The attoCSFM has optical access for the confocal part, including a standard objective holder set onto a slip-stick positioner in $z$ direction for the focus.
The AFM part is a combination of two positioning and scanning towers.
Each tower has an $xyz$ positioning stack of slip stick positioners with \SI{15}{\milli\meter} travelling range and an attached $xyz$ piezo open-loop scanner with \SI[product-units = power]{21x21}{\micro\meter} scan range in $x$ and $y$ and \SI{7}{\micro\meter} in $z$.
One tower is used to hold the diamond sample, while the other is scanning the AFM tip.
The AFM's interferometric readout of the cantilever position was replaced with a tuning-fork based readout system, which allows scanning in non-contact mode.
In addition, the tuning-fork signal was converted using a home-built transimpedance amplifier, realized with an operational amplifier following the approach of Ref.~\onlinecite{Jahncke_2004}.
A third $xyz$ positioning stack is used to position and align a permanent magnet for the creation of the external bias field.
The microscope is encased in a housing that is airtight and can support exchange atmosphere or a medium vacuum if desired.
The microscope is computer controlled, mostly via a self written software, developed using the Python programming language.\footnote{\href{https://www.python.org}{Python Software Foundation}} Full access to the source code allows a flexible working environment with a high degree of automation.

\subsection{External Magnetic Field}\label{magnet}

As mentioned earlier, a large magnetic bias field and proper alignment of said field is of central importance to establish efficient SSR.

In the presented setup we use a permanent magnet manufactured from NdFeB, since they are commercially available \footnote{\href{https://www.hkcm.de/HKCM_select.php?fav=62151id&l=de}{HKCM Engineering e.K.}} and can generate remanent fields as high as $\SI{1.3}{\tesla}$.
Also, such a magnet does not introduce any additional heat load due to the presence of an induction current, as in the case for electromagnets.

The geometry of the magnet was designed carefully in order to reflect the following aspects.
We employ a (100) diamond sample, i.e.\ the NV center's $z$-axis is tilted relative to the diamond surface by \SI{54.75}{\degree}, along which the magnetic field needs to be aligned.
The magnet has to fit into the device and leave enough space above the diamond sample to accommodate the tuning fork AFM head for scanning probe measurements.
The gradient of the magnetic field should be small both to facilitate alignment and to be less sensitive to drift.
At the same time the absolute magnetic field at the site of the NV center should be as large as possible.

\begin{figure}[!htbp]
	\centering
		\includegraphics[width=1.0\columnwidth]{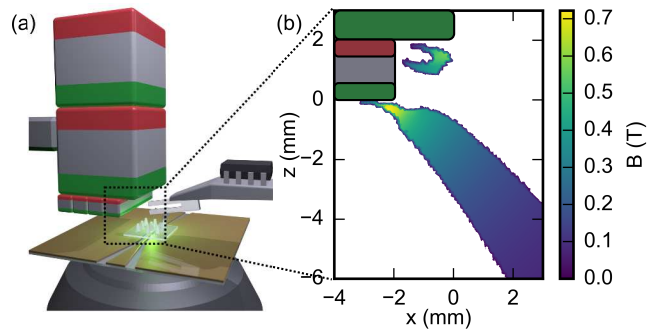}
  \caption{Strong permanent magnet in a scanning probe microscope setup. (a) In the middle of the picture is the diamond membrane with nanopillars (everything except the nanopillars is to scale) resting on the microwave stripline above the objective. The tuning-fork holder with preamplifier circuit can be seen on the right. The magnet formed by two \SI{1}{\centi\meter} cube permanent magnets and four smaller rods is supported and approaches the diamond from the left. (b) Close-up of a $z$-$x$ cross-section of a finite-element analysis for the stray magnetic field from the magnet displayed in (a), area indicated by dotted rectangle. In the upper left corner the very end of the magnet is indicated.
	The magnetic field strength is only plotted in regions, where the alignment on the NV center's axis is better than \SI{99}{\percent}.
	With this magnet design aligned fields as high as \SI{0.7}{\tesla} can be achieved.}
	\label{fig:magnet}
\end{figure}

The final configuration of the magnet is displayed schematically in Fig.\,\ref{fig:magnet}a, together with the diamond sample membrane and the tuning fork used in scanning probe.
The magnet is put together from two \SI[product-units = power]{10 x 10 x 10}{\milli\meter} cubes and four \SI[product-units = power]{10 x 2 x 2}{\milli\meter} rods.
This design was chosen according to a finite-element-analysis of the stray magnetic field, shown in Fig.\,\ref{fig:magnet}b.

In order to align the magnetic field, the permanent magnet is mounted onto an $xyz$ slip stick positioner from attocube.
The first step in alignment is to scan the magnet in the $xy$-plane above the NV center while recording the fluorescence.
Misalignment affects the electron spin as it causes mixing of the electron spin sub-levels of the NV center and thus reduces the fluorescence count-rate.\cite{Tetienne2012}
After this coarse alignment, the $xy$-scan is repeated in a focus area of highest fluorescence, with the SSR fidelity being evaluated at every pixel.
The maximum is then chosen as the final position of the magnet (see Fig.\,\ref{fig:alignment}b).
The absolute strength of the magnetic field is then determined by the distance between the magnet and the diamond sample.
With the additional constraint of the $xy$-alignment, this means the previously chosen $z$-value defines the final magnetic field strength.
This way, we can reach values of up to \SI{0.7}{\tesla}.

Once the magnet is brought into proper position and provided that the system has a good temperature stability, the magnetic field strength and alignment are remarkably stable (see Fig.~\ref{fig:stability}).
In the next section we show how the necessary stability of our system was realized.

\subsection{Temperature Stabilization}\label{temp_stability}

Drift poses a problem for most cutting-edge microscopy systems, especially at ambient conditions.
Our system combines two microscopy techniques, namely confocal and scanning probe microscopy.
The as-purchased instrument is constructed exclusively from titanium, which is non-magnetic and features a relatively low thermal expansion coefficient of around \SI{8.6e-6}{\per\kelvin}. \cite{Hidnert1943}
It is further stabilized by keeping the inside of the microscope at a slightly elevated temperature relative to the environment via a proportional-integral (PI) feedback loop.
This way drift-rates of less than \SI[per-mode=symbol]{5}{\nano\meter\per\hour} measured by the AFM were achieved, which is good enough from the microscopy point of view.

\begin{figure}[!htbp]
\centering
		\includegraphics[width=1.0\columnwidth]{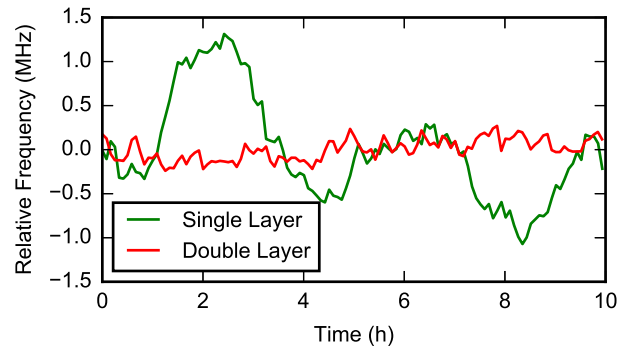}
  \caption{Drift of the NV center's transition frequency as a function of time with single layer (green) and double layer (red) thermal stabilization. With the latter, the frequency shifts are reduced to about \SI{0.5}{\mega\hertz} peak-to-peak in the course of \SI{10}{\hour}.}
	\label{fig:stability}
\end{figure}

However, the stability of the transition frequency of the NV center is not sufficient for SSR, where the two hyperfine lines of the $^{15}\mathrm{N}$ nucleus need to be addressed individually (see section~\ref{SSR_implementation} and Fig.\,\ref{fig:sequence}b).
This means that the peak-to-peak frequency drift needs to be smaller than the full width at half maximum (FWHM) of the hyperfine resolving ODMR line of about \SI{.7}{\mega\hertz}, which was originally not the case (see Fig.\,\ref{fig:stability}, green line).
It is possible to compensate for the frequency drifts by frequent ODMR measurements, but these take time and thereby reduce the achieved SNR.

The frequency drift is caused by a change in magnetic field strength at the site of the NV center, which can be due to a drift of the relative position or due to thermally induced changes of the magnetization of the permanent magnet.
From the finite-element-analysis simulation, a magnet position yielding a field of \SI{398}{\milli\tesla} (value used in experiments)  corresponds to a magnetic field gradient of \SI[per-mode=symbol]{0.15}{\milli\tesla\per\micro\meter}.
The above mentioned maximum value of \SI{0.7}{\mega\hertz} in frequency drift would then translate into a necessary temperature stability of \SI{0.39}{\kelvin},\footnote{Based on \SI{0.7}{MHz} corresponding to \SI{166}{\nano\meter} position change, which corresponds to a \SI{2.84e-6}{} relative change in the \SI{5}{\centi\meter} titanium rod holding the magnet.}
which is easily realized.
The temperature stability of the magnetization, however, is given by \SI[per-mode=symbol]{0.11}{\percent\per\kelvin}. \cite{Ma2002}
This value translates into a necessary temperature stability of \SI{0.06}{\kelvin} peak-to-peak for a frequency stability better than \SI{.7}{\mega\hertz}.

The most straightforward way to achieve this high level of stability was to add a second temperature stabilization layer. 
The microscope was encased in a styrofoam box of a thickness of \SI{60}{\milli\meter}, with a lining of acoustical foam to simultaneously reduce acoustic coupling to the scanning probe system.
The breadboard, which carries the microscope inside this second layer, was temperature stabilized by a second PI loop.
The setpoint for the breadboard temperature was chosen to lie between the setpoint of the inside of the microscope and the ambient temperature in the laboratory.\footnote{The temperature stability in the laboratory itself is approximately \SI{0.5}{\kelvin} peak-to-peak.}

With both layers of the temperature control active, the stability in the magnetic field and therefore the stability of the NV center's ODMR transition frequencies is improved to fluctuations of about \SI{0.5}{\mega\hertz} peak-to-peak in the course of \SI{10}{\hour}, as demonstrated in Fig.\,\ref{fig:stability} (corresponding to \SI{18}{\micro\tesla} or temperature changes of the magnet of approximately \SI{0.04}{\kelvin}).

\subsection{Spin manipulation setup}
The NV center's electron spin is manipulated via microwave (MW) radiation in the \si{\giga\hertz} regime, while the $^{15}\mathrm{N}$ nuclear spin is manipulated with radiofrequency (RF) fields in the \si{\mega\hertz} regime ($\gamma_{^{15}\mathrm{N}} = \SI[per-mode=symbol]{-4.3}{\mega\hertz\per\tesla}$, $\gamma_{\mathrm{e}^-} = \SI[per-mode=symbol]{28}{\giga\hertz\per\tesla}$).

\begin{figure}[!htbp]
	\centering
	\scriptsize
	\includegraphics[width = 1.0\columnwidth]{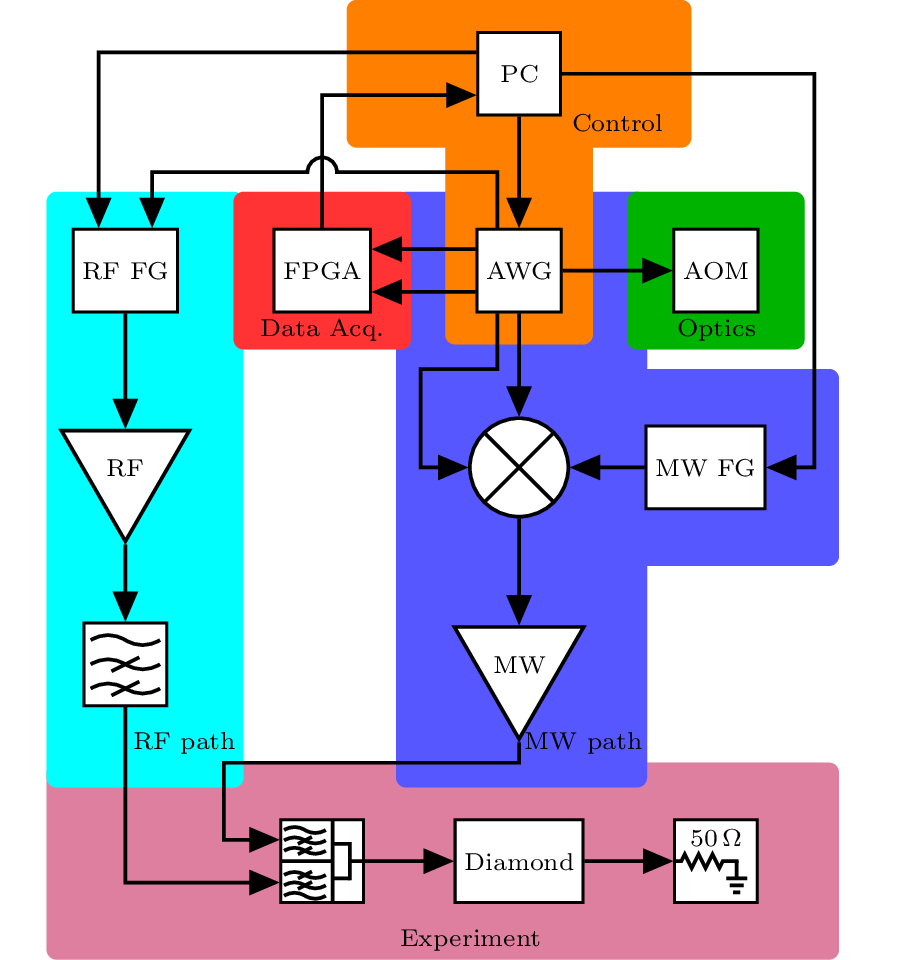}
	\caption{Diagram of the spin manipulation setup, separated into five subcategories. Control (orange), Radio-Frequency path (cyan), Data Acquisition (red), Microwave path (blue), Optics (green) and finally the actual Experiment at the diamond sample (purple). For a detailed description see main text.}
	\label{MW_setup_diagram}
\end{figure}

The spin manipulation setup can be separated into five categories as color-indicated in Fig.\,\ref{MW_setup_diagram}.
The control part (orange) is comprised of the primary control PC, which is used to set the frequencies of the RF and MW sources and analyze the readout data from a field-programmable gate array (FPGA) used for data acquisition (red).
Additionally, it feeds the Arbitrary Waveform Generator (AWG) (Tektronix AWG520) with the desired measurement pulse sequence.

The AWG falls into two categories, on the one hand it is the first stage of the MW path (blue) which synthesizes the MW pulses.
On the other hand the AWG is the central device that synchronizes the different paths of the measurement system.
For this purpose it is sending triggers to the RF function generator in the RF path (cyan), to the data acquisition FPGA and to the optics via the acousto-optic modulator (AOM) (green) to pulse the laser.
The optics path itself consists of a standard confocal setup.

The MW path is comprised of the AWG giving out two \SI{90}{\degree} phase shifted waveforms in the \SI{100}{\mega\hertz} range, which are fed into an IQ mixer (Marki MLIQ-0218L) in single sideband modulation mode, together with the high frequency carrier in the \si{\giga\hertz} range, which is generated by a R\&S~SMR20 signal generator.
The modulated high frequency signal is finally sent to a TriQuint~RM022020~Spatium amplifier.

The RF path begins with a Rigol~DG1032Z as the RF function generator, which is gated by the AWG.
The RF pulses are then amplified with a Mini-Circuits~LZY-22+ amplifier.
Subsequently they are sent through a home built \SI{1}{\mega\hertz} high pass filter to prevent previously observed low frequency ringing.

Finally, the RF signal is combined with the MW signal via an AMTI Microwave Circuits D1G018G3 diplexer before being sent to the diamond sample via a coplanar waveguide stripline (purple).
To prevent reflections of the transmitted power, the system is terminated with \SI{50}{\ohm}.

Data acquisition is done via a TimeTagger \footnote{\href{http://swabianinstruments.com/timetagger.html}{Swabian Instruments Time Tagger 20}} based on an FPGA, which receives the photon clicks from the avalanche photodiodes (APDs) and all necessary triggers from the AWG to bin them accordingly in a time-resolved fashion.
This data is then provided to the computer (PC) for analysis.

\subsection{SSR Implementation}\label{SSR_implementation}

\begin{figure}[!htbp]
		\centering
		\includegraphics[width=0.50\textwidth]{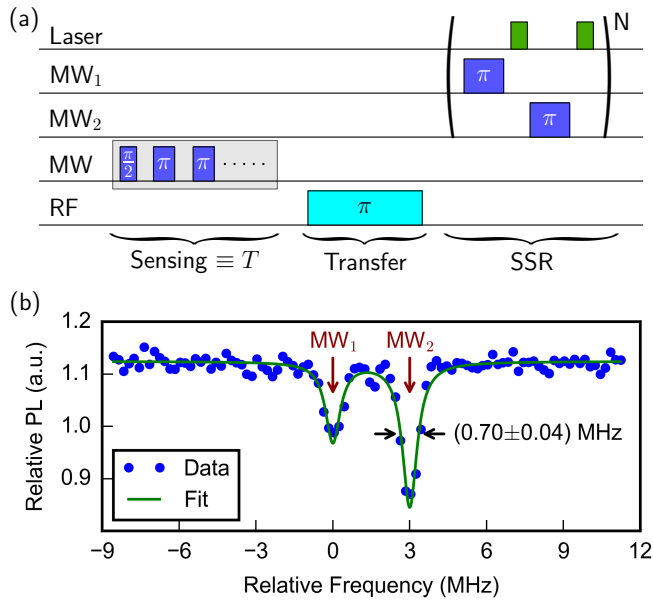}
  \caption{(a) SSR assisted measurement scheme. The measurement is divided into three parts, consisting of the \textsf{Sensing} part, conducted on the electron spin during the sensing time $T$ using strong, broad microwave pulses (MW). This is followed by the \textsf{Transfer} of the electron spin state onto the nuclear spin via a CNOT gate realized by a radio-frequency (RF) pulse on the nuclear spin.
	Finally the \textsf{SSR} of the nuclear spin is conducted via $N$ repetitions of CNOT gates on the electron spin (soft, therefore narrow MW$_1$ and MW$_2$ pulses, see also (b)) with subsequent readout laser pulse. The readout is conducted on both $^{15}\mathrm{N}$ transitions, for robustness reasons.
	(b) Typical ODMR of the $\ket{m_s=0} \rightarrow \ket{m_s=-1}$ transition showing the hyperfine splitting due to the $^{15}\mathrm{N}$ nucleus. The two probing frequencies MW$_1$ and MW$_2$ are indicated.}
	\label{fig:sequence}
\end{figure}

With the capability to manipulate both the electron spin and the nuclear spin, it is now possible to implement SSR-assisted measurements.
Figure~\ref{fig:sequence} depicts such a measurement sequence, which consists of three parts:

\begin{enumerate}
	\item During the sensing part of length $T$, an arbitrary measurement sequence is run on the NV center's electron spin (e.g. a Hahn echo sequence or a relaxometry sequence.\cite{Schmid-Lorch2015}).
	\item In a non-SSR assisted measurement this would be followed by optical readout and destruction of the electron spin state.
		
	In the SSR-assisted case the spin information is instead transferred onto the nuclear spin via a controlled NOT (C$_\mathrm{e}$NOT$_\mathrm{n}$) gate in the form of an electron spin selective $\pi$-pulse on the nuclear spin.\footnote{For the applied field of \SI{398}{\milli\tesla} the $\pi$-pulse length on the $^{15}\mathrm{N}$ was measured to be \SI{35.5}{\micro\second} at a resonance frequency of \SI{1.717}{\mega\hertz}.}
	\item Finally, the nuclear spin state is read out via the actual SSR.
\end{enumerate}

The readout of the nuclear spin state via SSR is realized with a C$_\mathrm{n}$NOT$_\mathrm{e}$ gate.
Such a gate is a nuclear spin state selective $\pi$ pulse - it flips the electron spin conditional on the nuclear spin state.
The C$_\mathrm{n}$NOT$_\mathrm{e}$ gate is then followed by a laser pulse to read out and repolarize the electron spin.
The recorded count-rate is reduced if the nuclear spin resides in the probed state, and left untouched otherwise. 
This change in count-rate can be used to determine the nuclear spin state.\cite{Neumann2010b}
However, the readout relies heavily on a stable underlying count-rate, which is not always the case.
Specifically, with surface-sensitive shallow NV centers the count-rate tends to fluctuate and change over time. \cite{Bradac2010}
To circumvent this problem, we probe both nuclear spin states virtually in a simultaenous manner by alternating the MW frequency between the two hyperfine lines ($\mathrm{MW}_1$ and $\mathrm{MW}_2$ in Fig.\,\ref{fig:sequence}).
This readout is repeated $N$ times with the photon numbers being counted separately for the two transitions and finally subtracted from each other.
The sign of the result reflects the nuclear spin state.
Since fluctuations in the count-rate affect both transitions in the same way, they will cancel out during the subtraction, making the readout more robust.

\section{Performance}\label{performance}

All of the following measurements were conducted at a magnetic field of $\SI{398}{\milli\tesla}$, which corresponds to a transition frequency between the $\ket{m_s=0}$ and $\ket{m_s=-1}$ sub levels of $\SI{8.274}{\giga\hertz}$.
The magnetic field was aligned prior to the measurements via fluorescence and SSR fidelity, as described in section~\ref{magnet}.

\subsection{Enhanced Photon Collection with Nanopillars}

A single NV center in a standard diamond membrane with shallow implantation yields in our case a count-rate of around $\SI{150}{\kilo\hertz}$.\footnote{All count-rates reported in this article were measured at saturation.}

We compare this to the count-rate from a single NV center situated inside a tapered nanopillar, which showed a photon flux of more than $\SI{760}{\kilo\hertz}$. This is an improvement of a factor of about $5$ in count-rate, which corresponds to a two-fold improvement in the SNR.

\subsection{Single Shot Readout}\label{performance_SSR}

The MW power applied during the C$_\mathrm{n}$NOT$_\mathrm{e}$ was chosen to result in a $\pi$ pulse length of about \SI{1}{\micro\second}, which is a good value to have a high selectivity of the gate within a reasonable time.
The length of the laser pulses for the SSR was chosen to be \SI{200}{\nano\second} following Ref.~\onlinecite{Neumann2010b}, since it is a good trade-off between high contrast in electron spin readout and re-polarization rate.
Additionally, a short laser illumination is also of benefit concerning the nuclear spin lifetime, because the lifetime decreases under laser illumination.

\begin{figure}[!htbp]
\centering
	\scriptsize
	\includegraphics[width = 1.0\columnwidth]{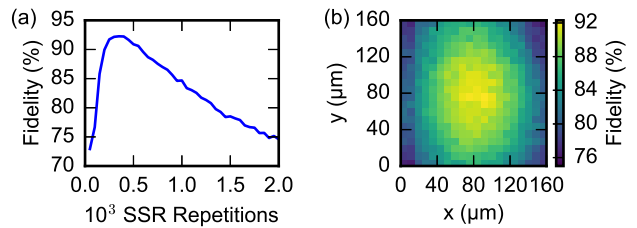}
	\caption{Optimization of single-shot readout scheme. (a) Fidelity as a function of the number of repetitions (laser pulses) in an SSR pulse sequence. The maximum fidelity achieved in the setup was \SI{92}{\percent}. (b) Magnetic field alignment using the SSR fidelity.}
	\label{fig:alignment}
\end{figure}

The number of SSR repetitions, $N$, was optimized experimentally by recording the SSR fidelity while sweeping the repetitions.
Figure~\ref{fig:alignment}a displays the SSR fidelity as a function of SSR repetitions with a maximum of $\mathcal{F}=\SI{92}{\percent}$ at about $N=400$.

With the optimal number of repetitions, a second, finer magnetic field alignment based on SSR fidelity was performed (see Fig.\,\ref{fig:alignment}b).
Once aligned, small frequency deviations can be corrected by ``refocusing'', i.e.\ measuring the ODMR frequency again.

\subsection{Benchmark Experiment}

For benchmark purposes, we compare measurements with and without the aid of SSR.
Since SSR works in principle with any measurement scheme conducted on the NV center, we chose the relaxometry or $T_1$ measurement scheme, which is simple to implement and interpret.
Additionally it is operating well within the regime of long sensing times, where the SSR can demonstrate its capabilities.

The $T_1$ decay was measured by letting the polarized electron spin of the NV center evolve during different free evolution times, $\tau$, followed by the readout.
In our example the values for $\tau$ were chosen to range from $\SI{1}{\micro\second}$ up to $\SI{50}{\milli\second}$ with a logarithmic spacing.
Every value is measured twice, with the second measurement being followed by a MW $\pi$ pulse, which inverts the spin state.
This inversion is used as a reference measurement to compensate for charge state effects and count rate fluctuations.\cite{HaeberleT.2015}

The full measurement scheme is conducted twice, once with the standard readout (simple laser pulse) and once with SSR (composed of the RF pulse and the actual SSR).
To compare the measurements, each run was integrated for the same amount of time.

\begin{figure}[!htbp]
		\centering
		\includegraphics[width=1.0\columnwidth]{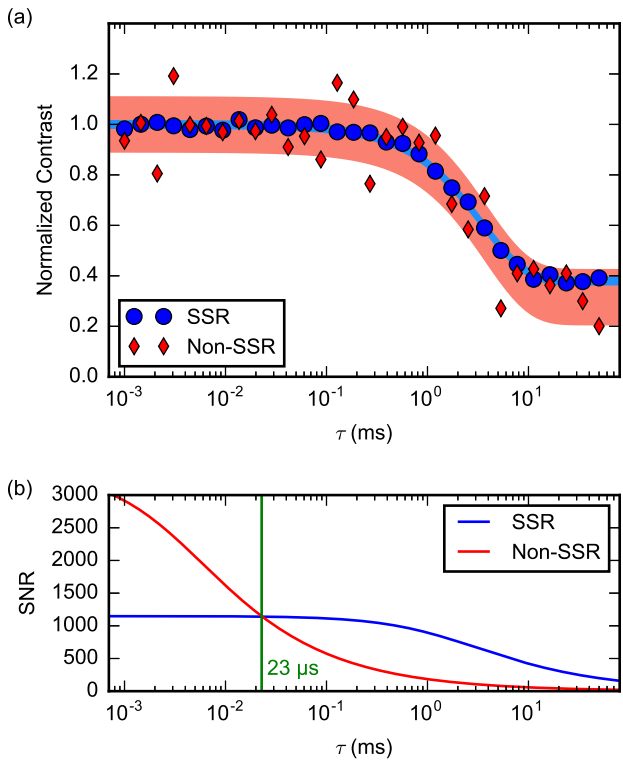}
  \caption{(a) Comparison between a $T_1$ measurement with SSR (blue) and with standard readout (red) for an integration time of \SI{1}{\hour}. Both datasets were fitted with an exponential function. The standard deviation between the fit and the data was determined and is plotted as the lightly shaded uncertainty regions, standard readout, red, SSR, blue.
	(b) Comparison of SNR as a function of $\tau$ for the experimental parameters, derived from theory. The break-even point where SSR becomes more efficient is reached for $\tau=\SI{23}{\micro\second}$ (green line).}
	\label{fig:compare3600}
\end{figure}

Figure~\ref{fig:compare3600} displays the results for an integration time of $\SI{1}{\hour}$.
The non-SSR measurement managed to complete 11200 sweeps during the integration time, the SSR measurement only 8690.
However, since the SSR is much more efficient, the result in Fig.\,\ref{fig:compare3600} clearly shows an improved SNR for SSR compared to standard readout.
The results were fitted with an exponential function to determine the characteristic $T_1$ decay time and the SNR of the measurement in a quantitative way.

Table~\ref{tab:results3600} summarizes the results from the two measurements.
The estimated $T_1$ time is practically the same, especially if one takes the errors into account.
The more interesting aspect is the difference in SNR for the two measurements, which is a factor of 8.6 larger in the SSR measurement.
This is also in a reasonable agreement with theory (see Sections~\ref{SNRofStandard} and \ref{SNRofSSR}).
The SNR of the measurement is even slightly higher than what the theory predicts.

\begin{table}[!htbp]
	\centering
	\caption{Results from the $T_1$ measurement data analysis for an integration time of \SI{1}{\hour}. Both measurement schemes yield consistent $T_1$ times, but the SSR aided scheme has a factor 8.6 higher SNR. This also fits reasonably well to the estimate from theory, both in the individual SNR values and in the improvement factor of 6.8.}
		\begin{ruledtabular}
		\begin{tabular}{lcc}
					& Non-SSR & SSR \\
		\hline
		$T_1$	& $\SI{3.8(10)}{\milli\second}$ & $\SI{3.5(1)}{\milli\second}$ \\
    SNR		& 9.4 & 80.5 \\
		SNR from Theory		& 10.3 & 70.3 \\
		\end{tabular}
		\end{ruledtabular}
	\label{tab:results3600}
\end{table}

As previously mentioned, the SSR measurement is more efficient than the standard readout for longer sensing times, due to its overhead.
With our experimental parameters the break-even point can be assessed to a sensing time of $\tau=\SI{23}{\micro\second}$ (see Fig.\,\ref{fig:compare3600}b).
In our case, for any sensing time longer than that, SSR will perform better than the standard readout.

\subsection{Performance Summary}

In summary, a significant gain is obtained both via the SSR-assisted readout and the tapered nanopillars.
From the nanopillars we get an improvement of the SNR by a factor of 2.3.
In the exemplary $T_1$ measurement we showed a further improvement via SSR of a factor of 8.6.
In combination, this means that compared to a non-pillar sample operated with standard readout, we have an SNR which is a factor of 19.3 higher.
In terms of acquisition time this corresponds to a reduction of a factor of 373 to reach the same SNR.
In other words, the one hour integrated SSR measurement shown in Fig.\,\ref{fig:compare3600} would have taken more than 15 days on a non-enhanced setup to reach the same SNR.
We would like to point out, that the two improvements are not fully independent of each other, as the SSR strongly benefits from a high count-rate, which allows a higher fidelity in the first place.

\section{Outlook}

The tremendous reduction in acquisition time will make previously time-consuming measurements possible. This is particularly significant in scanning probe applications, where the integration time per pixel reaches up to minutes,\cite{RugarD.2014, DeVience2015, HaeberleT.2015, Pelliccione2014, Schmid-Lorch2015} rendering high-resolution scans practically impossible.
Additionally, more information can be acquired in the same time, e.g. full spectra at each pixel instead of just a few data points.

The ability to engage SSR and manipulate nuclear spins along with the NV center's electron spin, grants access to a very interesting hybrid quantum system between sensitive sensor and long-lived memory.
This allows cutting edge quantum metrology applications, such as nuclear magnetic resonance (NMR) with a vastly enhanced magnetic field sensitivity as has been recently shown in Ref.~\onlinecite{Zaiser2016}.
These results were extended to other nuclear spins with further improved NMR resolving even chemical shifts, as shown in Ref.~\onlinecite{Aslam2016}.

With a clear set of guidelines to estimate from which sensing times it is preferable to use SSR, the presented system is now in a state where these techniques can be applied in scanning probe microscopy.

\section{Acknowledgements}

We would like to thank Johannes Greiner, Florestan Ziem and Nabeel Aslam for fruitful discussions. We acknowledge support from the EU FP7, grant No.~611143 (DIADEMS). F.F.O.\ acknowledges CNPq for the financial support through the project No.~204246/2013-0. A.F.\ acknowledges support from the Alexander von Humboldt Foundation.

\section*{Appendices}\label{appendices}

\subsection{SNR of standard readout}\label{SNRofStandard}

The SNR of the standard measurement is defined by shot noise.
To estimate the SNR for given parameters, we are looking at the fluorescence answer of the readout laser pulse.
The fluorescence answer of the readout laser pulse is divided into a detection and a reference window, where the detection window is normalized to the reference window.
This way the measurement becomes more robust against count-rate fluctuations.

The actual signal is then found by repeating the measurement with a final $\pi$ pulse at the end to invert the electron spin state.
Thereby, additional artefacts can be prevented, e.g.\ from a transition to the neutral charge state $\mathrm{NV}^0$.

With the two detection window counts, $n_{(0,\pi)}$, and the respective reference window counts, $n_{(0,\pi),r}$, the signal, $S_\mathrm{std}$, is given by:
\begin{equation*}
	S_\mathrm{std} = \frac{n_0}{n_{0,r}}-\frac{n_{\pi}}{n_{\pi,r}}.
\end{equation*}
The error, $\Delta S_\mathrm{std}$ can then be calculated by:
\begin{align}
	\Delta S_\mathrm{std} & = \left[\left(\frac{1}{n_{0,r}}\Delta n_0\right)^2 +\left(\frac{1}{n_{\pi,r}}\Delta n_\pi\right)^2 +\right. \nonumber\\
	& \qquad \left.\left(\frac{n_0}{n_{0,r}^2}\Delta n_{0,r}\right)^2+ \left(\frac{n_\pi}{n_{\pi,r}^2}\Delta n_{\pi,r}\right)^2\right]^{1/2}.
\label{eq:standard_primary_error}
\end{align}
This expression can be simplified, since several of the variables above have the same values, at least to a first approximation.
The photon numbers can be simplified to
\begin{align*}
	n_{0,r}&=n_{\pi,r}=n \\
	n_0&=c_0 n \\
	n_{\pi}&=c_\pi n,
\end{align*}
and the errors by
\begin{align*}
	\Delta n_{0,r}&=\Delta n_{\pi,r}=\Delta n = \sqrt{n} \\
	\Delta n_0&=c_0 \Delta n \\
	\Delta n_{\pi}&=c_\pi \Delta n.
\end{align*}
Here, $c_0$ is the count-rate enhancement of the $\ket{m_s=0}$ state and $c_{\pi}$ is the enhancement of the $\ket{m_s = -1}$ state. \cite{PhysRevB.81.035205}
Inserting this into equation~\eqref{eq:standard_primary_error} yields
\begin{equation*}
	\Delta S_\mathrm{std} = \sqrt{\frac{2}{n} \left(c_0^2 + c_\pi^2\right)}.
\end{equation*}
The SNR is given by $\frac{S_\mathrm{std}}{\Delta S_\mathrm{std}}$:
\begin{equation*}
	\mathrm{SNR}_\mathrm{std} = \sqrt{\frac{n}{2}}\frac{c_0 - c_\pi}{\sqrt{c_0^2 + c_\pi^2}}.
\end{equation*}

This calculation was also conducted with taking background fluorescence into account. However, the difference in the results was practically non-existent, which is why it was left out to keep the calculation clearer and easier to follow.

We can now insert the numbers from the measurement displayed in Fig.\,\ref{fig:compare3600}.
There we had $c_0=1.12$ and $c_\pi=0.73$, $n$ can be calculated from the number of sweeps $11200$, the length of the acquisition window \SI{300}{\nano\second} and the count-rate \SI{760}{\kilo\hertz}.

This, then, leads to a theoretical value of $\left.\mathrm{SNR}_\mathrm{std}\right|_\mathrm{1 hour} = 10.3$.

\subsection{SNR of SSR}\label{SNRofSSR}

SSR is a projective measurement, where every single measurement yields either nuclear spin up or down, say 0 or 1, respectively.
The measurement is therefore following a binomial distribution.
Once again, two measurements, one with inverting $\pi$ pulse and one without are conducted.
The signal for the SSR is given by the difference of the spin non-flip probabilities for the measurement, $p_0$, and the inverted measurement $p_\pi$:
\begin{equation*}
	S_\mathrm{SSR} = p_0 - p_\pi.
\end{equation*}
The standard deviation for $N_\mathrm{sweeps}$ repetitions of the SSR is given by:
\begin{equation*}
	\sigma = \sqrt{\frac{p\left(1-p\right)}{N_\mathrm{sweeps}}}.
\end{equation*}
The error $\Delta S_\mathrm{SSR}$ can be derived as:
\begin{equation*}
	\Delta S_\mathrm{SSR} = \sqrt{\frac{1}{N_\mathrm{sweeps}}\left(p_0\left(1-p_0\right) + p_\pi\left(1-p_\pi\right)\right)}.
\end{equation*}
For the SNR this yields
\begin{equation*}
	\mathrm{SNR}_\mathrm{SSR} = \frac{S_\mathrm{SSR}}{\Delta S_\mathrm{SSR}} = \frac{\left(p_0-p_\pi\right)\sqrt{N_\mathrm{sweeps}}}{\sqrt{p_0\left(1-p_0\right) + p_\pi\left(1-p_\pi\right)}}.
\end{equation*}

In the measurement the values $p_0=0.85$ and $p_\pi=0.40$ were observed.
These values agree with Ref.~\onlinecite{Waldherr_2011}.
With $N_\mathrm{sweeps}=8690$, this leads to a theoretical prediction of $\left.\mathrm{SNR}_\mathrm{SSR}\right|_\mathrm{1 hour} = 70.3$.

The estimated improvement factor of the SSR measurement compared to the standard measurement is therefore 
\begin{equation*}
	\frac{\mathrm{SNR}_\mathrm{SSR}}{\mathrm{SNR}_\mathrm{std}} = 6.8.
\end{equation*}
This value is even slightly below the measured value of $8.6$.

Figure~\ref{fig:field} shows the SNR improvement factor $\mathrm{SNR}_\mathrm{SSR}/\mathrm{SNR}_\mathrm{std}$ dependent on the SSR fidelity and the sensing time, as it was derived from theory.
For the calculation of $\mathrm{SNR}_\mathrm{std}$ the same values were taken as in section~\ref{SNRofStandard}.
For $\mathrm{SNR}_\mathrm{SSR}$ the fidelity value was varied and the contrast was calculated based on the NV center residing $30\%$ of the time in the NV$^0$ state.\cite{Waldherr_2011}
For both the number of sweeps was calculated using the varied sensing time.

\subsection{Diamond Sample}
We use commercially available electronic-grade chemical vapor deposition (CVD) diamond. \footnote{\href{http://e6cvd.com/us/el-sc-plate-4-5x4-5-mm-0-5-mm-thick.html?___SID=U}{Element Six, item no.~145-500-0390}}
This diamond is then cut to a thin ($\sim\SI{30}{\micro\meter}$) membrane,\footnote{\href{http://usapplieddiamond.com/capabilities/laser-cutting/}{Applied Diamond, Inc.}} implanted at $\sim\SI{5}{\kilo\electronvolt}$ with $^{15}\mathrm{N}^+$ ions and annealed at a temperature of about \SI{950}{\celsius}.
Subsequently, an array of a few thousand nanopillars is fabricated using the same recipe as in Ref.~\onlinecite{Momenzadeh2015}.
This gives us the capability to select an NV center with the correct orientation with respect to the magnetic field (see Sec.~\ref{magnet}) while also having a $T_2$ (Hahn) of $\SI{108.1(14)}{\micro\second}$ and $T_1$ time of $\SI{3.5(1)}{\milli\second}$.

\subsection{Thermal Stability}

The temperature controller employed in our setup is a Lakeshore Model 335 Cryogenic Temperature Controller. It features two separate, PI controllable channels, each with a heating power of up to \SI{50}{\watt}.
The controller is configured and read out via a software written in the programming language Python.
This allows logging of the temperature and heating power development over time, which in turn can be used to determine when the system has reached stable operation and measurements can be commenced, usually after $0.5$~to~$\SI{1.5}\hour$.

\bibliography{ssr_rsi} 

\end{document}